\documentclass[letterpaper, 10pt, conference]{ieeeconf}
\usepackage[autostyle]{csquotes}
\usepackage{bm}
\usepackage{amssymb}
\usepackage{epsfig}
\usepackage{subfigure}
\usepackage{multicol, blindtext}
\usepackage{amsmath}
\usepackage{graphicx}
\usepackage{epstopdf}
\usepackage{xfrac}  
\usepackage{float}
\usepackage{algorithm}
\usepackage{algpseudocode}
\usepackage{amsfonts}
\usepackage{color}
\usepackage{dblfloatfix}
\usepackage{xfrac}  
\usepackage{multirow}%
\usepackage{fourier} 
\usepackage{array}
\usepackage{algorithm}
\usepackage{algpseudocode}
\usepackage{makecell}
\usepackage{blindtext}
\usepackage{subfiles} 

\providecommand{\U}[1]{\protect\rule{.1in}{.1in}}

\newtheorem{theorem}{\rm\textbf{Theorem}}

\newtheorem{remark}{\rm\textbf{Remark}}

\graphicspath{ {figures/} }
\IEEEoverridecommandlockouts

\begin{document}

\title{\LARGE\textbf {Optimal Merging Control of an Autonomous Vehicle in Mixed Traffic: an Optimal Index Policy } 
\thanks{This work was supported in part by NSF under grants ECCS-1931600,
DMS-1664644, CNS-1645681, CNS-2149511, by AFOSR under grant
FA9550-19-1-0158, by ARPA-E under grant DE-AR0001282, by the Math-
Works and by NPRP grant (12S-0228-190177) from the Qatar National
Research Fund, a member of the Qatar Foundation (the statements made
herein are solely the responsibility of the authors).
The authors are with the Division of Systems Engineering and Center for
Information and Systems Engineering, Boston University, Brookline, MA,
02446, USA, \texttt{{\small \{esabouni, cgc\}@bu.edu }}}}
\author{Ehsan Sabouni and Christos G. Cassandras}
\maketitle

\begin{abstract}
We consider the problem of a single Autonomous Vehicle (AV) merging into traffic consisting only of Human Driven Vehicles (HDVs) with the goal of minimizing both the travel time and energy consumption of the entire group of vehicles involved in the merging process. This is done by controlling only the AV and determining both the optimal merging sequence and the optimal AV trajectory associated with it. We derive an optimal index policy which prescribes the merging position of the AV within the group of HDVs. We also specify conditions under which the optimal index corresponds to the AV merging before all HDVs or after all HDVs, in which case no interaction of the AV with the HDVs is required. Simulation results are included to validate the optimal index policy and demonstrate cases where optimal merging can be achieved without requiring any explicit assumptions regarding human driving behavior.

\end{abstract}

\thispagestyle{empty} \pagestyle{empty}
\section{INTRODUCTION} 
The emergence of Connected and Automated Vehicles (CAVs) has the potential to drastically impact transportation systems in terms of increased safety, as well as reducing congestion, energy consumption, and air and noise pollution \cite{li2013survey}.
While infrastructure improvements typically offer only short-term solutions, 
research to date has shown that CAVs can provide long-term solutions to these problems through better information utilization and more precise trajectory design, especially in conflict areas such as intersections, roundabouts, and merging roadways all of which critically affect the performance of a traffic network \cite{MALIKOPOULOS_survey2017}.   
The main approach has been to formulate and solve optimal control problems for CAVs seeking to minimize travel times and fuel consumption while always satisfying safety constraints, e.g., \cite{ZHANG2019}.

The most important benefit of CAVs lies in the fact that they can \emph{cooperate} by sharing information and coordinating their respective motion. In contrast, a transportation system consisting entirely of Human Driven Vehicles (HDVs) operates based on the vehicle-centric (selfish) behavior of each driver who competes, rather than cooperating, with other drivers. This prevents a traffic network from achieving a much more efficient system-centric (socially optimal) equilibrium based on CAVs. To date, most of the research involving the optimal control of CAVs has been based on the assumption that all vehicles on a traffic network are CAVs and can cooperate with each other. Since such $100\%$ CAV penetration rate is likely to take a few decades before it can be realized \cite{ALESSANDRINI2015145}, it is crucial to investigate the ways in which the benefits of CAVs can be realized while they co-exist with HDVs in a \enquote{mixed traffic} environment. It is, for instance, possible that a partial presence of CAVs may even worsen the problem of congestion, since HDV behavior is stochastic and selfish by nature \cite{NMher2019}. It is therefore important to adapt the controllers or algorithms developed under the assumption of $100\%$ CAV penetration so that they can capitalize on the ability of CAVs to control their own motion in ways that minimize the unpredictability or selfish behavior of HDVs with which they interact. 

A common way to deal with CAVs in mixed traffic settings is to either make assumptions on the behavior of HDVs, such as maintaining constant speed in free flow traffic \cite{omidvar2020optimizing}, or adopting a specific car following model \cite{SUN2020102764},\cite{ZHAO201873}. Alternatively, the problem may be approached through data-driven control \cite{wang2021data} or learning-based algorithms to handle HDV stochasticity \cite{8569485}, \cite{8317694}. Despite some encouraging results, such trial-and-error methods are not yet applicable to real-time settings. 

A key question in a mixed traffic setting is ``what should the CAV penetration rate be to start seeing their beneficial effect on the existing transportation network?'' \cite{8511339},\cite{giammarino2020traffic}. Motivated by this question the goal of this paper is to investigate the extent to which a single CAV 
(which we will henceforth refer to as just an AV since it is not connected to any other vehicles)
can have a positive impact when interacting with a group of HDVs in terms of travel time and energy consumption for the HDVs and the AV itself. We consider the merging problem in such mixed traffic situations, specifically a group of $N$ HDVS approaching a merging point from one road while an AV approaches it from another road. The goal is to establish a ``building block'' for how a single AV can interact with HDV groups in more general traffic control settings where we can have two or more CAVs which can further enhance performance through connectivity.

Our contribution is to solve the problem of determining the optimal merging sequence in the above setting that benefits all vehicles in terms of both energy and time. In particular, the AV, indexed by $0$, can select one of $N+1$ possible ways to merge relative to the HDVs, i.e., $\{0,1,\ldots,N\}, \ldots, \{1,\ldots,N,0\}$. We provide an optimal trajectory for the AV to merge in the optimal order while guaranteeing safe merging. Moreover, we derive conditions under which the optimal merging sequence is either $\{0,1,\ldots,N\}$ or $\{1,\ldots,N,0\}$, i.e, the AV either proceeds ahead of or waits to merge behind the whole group. The importance of such a sequence is that it involves no disruption of the HDV group and requires no assumptions by the AV as to the behavior of the HDVs.

The paper is organized as follows. In Section II, we define the merging prblem framework and formulate the optimal control problem for minimizing both travel time and energy consumption of all $N+1$ vehicles. In Section III, the problem is reformulated so as to find the optimal merging sequence of the vehicles, as well as the associated optimal trajectories. Conditions for the two special sequences to be optimal are also derived. In Section IV, simulation results validate the general solution along with the special sequences that require no knowledge of HDV behavior. 


\section{Problem Formulation}
\label{sec:problem}
\begin{figure}[H] 
\vspace{-4mm}
\centering
\includegraphics[scale=1]{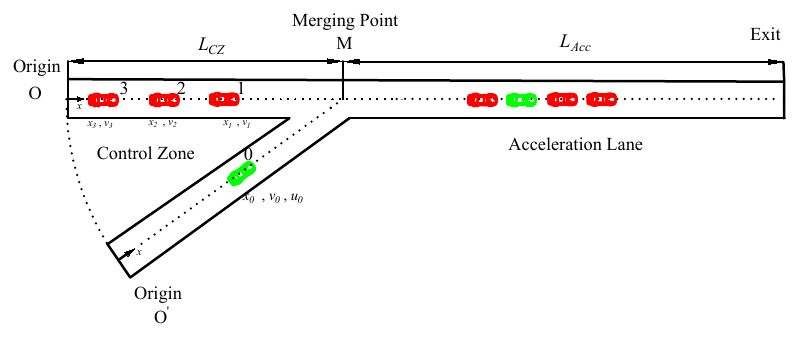} 
\caption{The merging problem }
\label{fig:merging}
\end{figure}

Similar to the merging problem studied in \cite{XIAO2021}, we consider traffic arriving from two roads joined at a Merging Point (MP) M where a collision might occur. In this paper, we consider situations where a single AV on one road merges with a group of HDVs coming from the other road as shown in Fig. \ref{fig:merging} by green and red vehicles, respectively. A \textit{Control Zone} (CZ) is defined to be an area within which vehicles can see other vehicles (if any) and obtain or possibly estimate their states using their on-board sensing equipment. The range of such zones, $L_{CZ}$ can vary for different vehicles as it depends on the geometry of the road and the vehicle sensing capabilities. Unlike the case of a CZ with $100\%$ CAVs and full communication capabilities, e.g., \cite{XIAO2021}, here HDVs do not have any connectivity. After crossing the MP, vehicles join a road segment called \textit{Acceleration Lane} (AL) of length $L_{AL}$, which is assumed to be long enough so that vehicles can reach their desired speed.

Throughout the paper, the index $0$ is reserved for the AV and let $t_0^a$ be the time when the AV arrives at the CZ from one road and detects any HDVs traveling on the other road. Then, $\mathcal{N}(t_0^a)=\{0,1,...,N(t_0^a)\}$ is the set of all (unique) indices in the CZ, where $N(t_0^a) \in \mathbb{N}$ is the total number of HDVs. 
If the AV detects an additional HDV entering its CZ, it is assigned the index $N(t_0^a)+1$ and added to $\mathcal{N}(t_0^a)$.
To ease notation, we will set $N=N(t_0^a)$ and $\mathcal{N}=\mathcal{N}(t_0^a)$ for the rest of the paper.

The AV dynamics are assumed to be of the form
\begin{equation} \label{VehicleDynamics}
\left[
\begin{array}
[c]{c}%
\dot{x}_{0}(t)\\
\dot{v}_{0}(t)
\end{array}
\right]  =\left[
\begin{array}
[c]{c}%
v_{0}(t)\\
u_{0}(t)
\end{array}
\right],  
\end{equation}
where $x_{0}(t)$ denotes the distance from the origin at which AV $0$ arrives, $v_{0}(t)$ denotes the velocity, and $u_{0}(t)$ denotes the control input (acceleration). The vehicle dynamics for the HDVs within the CZ are unknown to the AV, but we assume that their position $x_i(t)$ and velocity $v_i(t)$, $i \in \mathcal{N}-\{0\}$, can be directly obtained or estimated by the AV's on-board sensors.

The objective of the AV is to derive an optimal acceleration/deceleration profile so as to minimize the energy consumption and travel time of the total $N+1$ vehicles present in the CZ (including the AV itself). Therefore two objectives are defined as follows:

\textbf{Objective 1} (Minimize travel time): Let $t^f_i$ denotes the time vehicle $i$, $i \in \mathcal{N}$  leaves AL at the exit point. We wish to minimize the total travel time of all vehicles $\sum_{i=0}^{N}(t^f_i-t_0^a)$. It is important to note that the travel times of the HDVs are measured relative to the time that the AV observes them. Since HDVs cannot be controlled and there is no communication with them, their performance cannot be optimized without the presence of at least one AV which can decide when to reach the MP.

\textbf{Objective 2} (Minimize energy consumption): we wish to minimize the total energy consumption of all vehicles in the CZ,  $\sum_{i=0}^{N}E_i$, where the energy consumption model $E_i$ for each vehicle is expressed as:
\begin{equation} \label{energy}
E_{i}(t_{i}^{f},u_{i}(t))=\int_{t_{0}^{a}}^{t_{i}^{f}}\mathcal{C}(u_{i}(t))dt,
\end{equation}
where $\mathcal{C}(\cdot)$ is a strictly increasing function of its argument. For simplicity, we limit ourselves to the case where $\mathcal{C}(u_i(t))=\frac{1}{2}u_i^2(t)$. As with the way we defined the travel time for the HDVs, their energy consumption is also calculated starting at $t=t^a_0$.\\
\
We consider next the following constraints for the AV.

{\bf Constraint 1} (Safety constraints): Let $p \in \mathcal{N}-\{0\}$ denotes the index of
the HDV which physically immediately precedes the AV (if one is
present). We require that the distance $z_{0,p}(t):=x_{p}(t)-x_{0}(t)$
be constrained by:
\begin{equation}
z_{0,p}(t)\geq\varphi_c v_{0}(t)+\delta,\text{ \ }\forall t\in\lbrack
t_{0}^{a},t_{0}^{f}], \label{Safety}%
\end{equation}
where $t^f_0$ is the time that AV exits AL, it is worth mentioning that the safety constraint is only necessary for AL $(t>t_0^m)$ since there is only one AV on the side road as shown in Fig. \ref{fig:merging}. The reaction time for the AV here is denoted by $\varphi_c$ (as a rule, $\varphi_c=1.8s$ is used,
e.g., \cite{Vogel2003}) and $\delta$ is a given minimum safe distance.
 If we define $z_{0,p}$ to be the distance from
the center of the AV to the center of HDV $p$, then $\delta$ depends on the length of these two vehicles (generally dependent on the AV and HDV $p$ but taken to be a constant over all vehicles for simplicity).

{\bf Constraint 2} (Safe merging): Whenever the AV crosses a MP, there must be adequate safe space for the AV at this MP to avoid a possible lateral collision, i.e.,
\begin{equation}
\label{SafeMerging}
z_{0,j}(t_{0}^{m})\geq\varphi_c v_{0}(t_{0}^{m})+\delta,
\end{equation}
where $j$ is the index of the HDV that may collide with AV $0$ at the merging point. The choice of $j$ depends on the merging sequence that the AV determines which will be discussed in the next section. It is worth pointing out that HDV $j$ is also the  one that will precede the AV in the AL, therefore,  $j=p$ if such $p$ exists. Note that this constraint only applies at time $t_{0}^{m}$ and it provides a safe distance between AV and HDV $j$; however, in addition, the AV must ensure that the HDV $j+1$ (i.e. the HDV trailing the AV after merging) is willing and able to provide adequate space to avoid any collision. Thus, safe sets for the merging time and merging velocity will be derived to address this challenge which arises in mixed traffic scenarios.

{\bf Constraint 3} (Vehicle physical limitations): There are constraints
on the speed and acceleration of vehicles:
\begin{equation}
\begin{aligned} v_{min} \leq v_i(t)\leq v_{max}, ~\forall t\in[t_0^a,t_i^f]\end{aligned} \label{VehicleConstraints1}%
\end{equation}
\begin{equation}
\begin{aligned} u_{{min}}\leq u_i(t)\leq u_{{max}}, ~\forall t\in[t_0^a,t_i^f],\end{aligned} \label{VehicleConstraints2}%
\end{equation}
where $v_{max}> 0$, $v_{min} \geq 0$ denote the maximum and minimum speed allowed
for vehicles, and $u_{min}<0$, $u_{max}>0$ denote the minimum and maximum
control for vehicles, respectively.  \\
\textbf{Optimal Control Problem formulation.} Our goal is to determine a control law for the AV so as to achieve objectives 1-2 subject to constraints 1-3 for each $i \in \mathcal{N}$ and AV dynamics (\ref{VehicleDynamics}). We use the weight $\alpha\in[0,1]$ to construct a convex combination of time and energy metrics as follows:
\begin{equation} \label{eqn:tot_car_obj}
\begin{aligned}\min_{u_{0}(t),t_0^m} J(u_0(t),t_0^m)= \sum_{i=0}^{N}(\alpha(t_i^m-t_0^a)+(1-\alpha)E_i) \end{aligned} ~~t \in [t_0^a,t_0^m]
\end{equation}
subject to \eqref{VehicleDynamics}, \eqref{Safety}, \eqref{SafeMerging}, \eqref{VehicleConstraints1} and, \eqref{VehicleConstraints2}.
Note that the only decision variables here are the AV control input and AV merging time, since there is no control over the HDVs. The problem is formulated over $[t_0^a,t_0^m]$ since minimizing the objective function up to the MP implies the minimization beyond the MP as well. This problem is complicated by the fact that obtaining an expression for each HDV's travel time $t_i^m$ and energy consumption $E_i$, $i=1,\ldots,N$, as a function of the AV's states and control is a difficult task. To address this issue, we reformulate the optimal control problem so as to first derive an optimal merging sequence for all vehicles so as to minimize the objectives and subsequently determine the optimal control for the AV.

\section{OPTIMAL  INDEX  POLICY}
In this section, we derive an optimal index policy for the AV with the goal of minimizing the overall travel time and energy consumption of all vehicles as defined in \eqref{eqn:tot_car_obj}. In other words, we seek the optimal among all possible merging sequences of the AV and $N$ HDVs expressed as
\begin{equation} \label{J_k}
    \mathcal{K}=\Big\{ \{0,1,...,N\},\{1,0,...,N\},...,\{1,...,N,0\} \Big\},
\end{equation}
where, for example, the first sequence in $\mathcal{K}$ stands for the AV crossing the MP ahead of all $N$ HDVs.
In the sequel, we use the index $k\in \{1,\ldots,N+1\}$ to denote the $k$th element of the set $\mathcal{K}$. Thus, $k=1$ denotes the sequence in which the AV crosses the MP ahead of all $N$ HDVs and $k=N+1$ is the sequence where the AV is the last to cross the MP behind all HDVs. 
Therefore, we can rewrite the objective function in \eqref{eqn:tot_car_obj} in terms of its associated sequence $k\in \{1,\ldots,N+1\}$ as follows:
\begin{equation} \label{objective function general}
    J_k=\alpha \left( t^m_{0,k}-t_0^a+\sum_{i=1}^{N}t^m_{i,k} \right)   +(1-\alpha) \left(E_{0,k}+ \sum_{i=1}^{N}E_{i,k}  \right),
\end{equation}
    where $t^m_{i,k}$ and $E_{i,k}$ stand for the travel time and energy consumption (starting from $t_0^a$) of vehicle $i \in \mathcal{N}$, respectively, when the AV crosses the MP as the ${k}^{th}$ vehicle.  \\
\textbf{Assumption 1} : All HDVs have reached their desired speed $v_{i,k}(t)= v^d_i$, $t \in [t_0^a,t_0^f]$, $i \in \mathcal{N}-\{0\}$ and intend to pass the MP cruising at $v^d_i$ if not constrained by other road traffic.

Based on Assumption 1, the road leading to the CZ is long enough to ensure that each HDV has reached its steady-state velocity in the vicinity of MP. Moreover, this assumption implies that any deceleration by a HDV is perceived as a \enquote{disruption}. This allows us to express the travel time $t^m_{i,k}$ in (\ref{objective function general}) in terms of its minimal value and a disruption term $D^t_{i,k}$. 
Observe that the minimal value of $t^m_{i,k}$ corresponds to the index $k=N+1$, i.e., the AV crosses the MP after all $N$ HDVs, so we write
$t^m_{i,k}=t^m_{i,N+1}+D^t_{i,k}$, where $D^t_{i,k} \geq 0, \ \forall k \ \in \  \mathcal{K}-\{1,N+1\}$.

Similarly, for the energy consumption $E_{i,k}$ in (\ref{objective function general}), we write 
$E_{i,k}=E_{i,N+1}+D^E_{i,k}$ where $E_{i,N+1}$ is the amount of energy spent by HDV $i$ in the merging sequence $N+1$ where its disruption is zero, and $D^E_{i,k}$ is the excess energy that HDV $i$ spends because of a potential disruption in the merging sequence $k$. Thus, $D^E_{i,k} \geq 0, \ \forall k \ \in \  \mathcal{K}-\{1,N+1\}$. Further, since we have assumed in (\ref{energy}) that $\mathcal{C}(u_i(t))=\frac{1}{2}u_i^2(t)$, we have 
$E_{i,N+1}=0$, since there is no acceleration/deceleration in the HDV's undisrupted trajectory. Consequently, we get $E_{i,k}=D^E_{i,k}$.

As a result, by replacing $t^m_{i,k}$ and $E_{i,k}$ in \eqref{objective function general} by the disruption-based expressions above and by separating the AV's objective function $J_{A_k}$ from the objective function $J_{H_k}$ of the $N$ HDVs, \eqref{objective function general} for $k \in \mathcal{K}$ can be rewritten as:

\vspace{-2mm}
\begin{align} \label{objective function}
    J_k= &\underbrace{\alpha  (t^m_{0,k}-t_0^a)  + (1-\alpha) E_{0,k}}_{J_{A_k}} \\ \nonumber
    + &\underbrace{ \sum_{i=1}^{N}\alpha \bigl(t^m_{i,N+1}+D^t_{i,k} \bigl) +(1-\alpha) D^E_{i,k}}_{J_{H_k}}
\end{align}
Next, we evaluate each term in \eqref{objective function} starting with the disruption-based terms in $J_{H_k}$ pertaining to the HDVs. 

\textbf{HDV Objective Function $(J_{H_k})$ Evaluation :} To obtain an expression for $J_{H_k}$ we need to evaluate $t^m_{i,N+1}$, $D^t_{i,k}$ and, $D^E_{i,k}$ for $k \in \mathcal{K}$. The undisrupted travel time $t^m_{i,N+1}$ under Assumption 1 is given by:
\begin{equation}
    t^m_{i,N+1} = [L_{CZ}-x_i(t_0^a)] / v^d_i,
\end{equation}
since HDV $i$ cruises at $v^d_i$ until it crosses the MP.

The values of $D^t_{i,k}$ and $D^E_{i,k}$ depend on the $i$th HDV's behavior for all $i \geq k$ which are disrupted by the AV in merging sequence $k$.
To approximate these values, we approximate the disruption of HDV $k$, the first HDV that provides space for the AV to merge ahead of it and cross the MP, and then we use a discount factor $\gamma_{i,k} \in (0,1)$ to propagate the effect of the AV through the rest of the HDVs $i>k$. 
Let us assume that any HDV acceleration or deceleration rate is constant and given by $|\Bar{u}|$. 
Given the speed of the AV at the MP, $v_0(t^m_{0,k})$, and that of HDV $k$, $v_k^{d}$, and in view of Assumption 1, we can estimate $D^t_{i,k}$ and $D^E_{i,k}$ based on the difference between the HDV's desired velocity and the AV merging velocity and the HDV's deceleration $-\Bar{u}$ to provide enough space for merging. In particular, we can derive the following expressions for $D^t_{i,k}$ and $D^E_{i,k}$ for $i=k$ as follows (details are omitted but can be found in [Appendix.A]):
\begin{align}
    D^t_{k,k}(v_0(t^m_{0,k}))=\frac{\max(v^d_k-v_0(t^m_{0,k}),0)^2}{2\Bar{u}v_k^d}\nonumber\\
    D^E_{k,k}(v_0(t^m_{0,k})) = \frac{1}{2}\Bar{u} \max(v^d_k-v_0(t^m_{0,k}),0).
\end{align}
Note that the only controllable variable in these disruption measures is the AV's merging velocity $v_0(t^m_{0,k})$ which will be determined in the sequel. Now $D^t_{k,k}$ and $D^E_{k,k}$ can be used as the basis to calculate the disruption over all HDVs $i > k$ as follows:
\begin{eqnarray}\small
D^t_{i,k}(v_0(t^m_{0,k}))=\begin{cases} \label{epsilon}
 \gamma_{i,k} \Big[\frac{\max(v^d_k-v_0(t^m_{0,k}),0)^2}{2\Bar{u}v^d_k}\Big]  \ \ \ \ \ \  i > k\\
\\
0\ \ \ \ \ \  \textnormal{otherwise},
\end{cases}
\end{eqnarray}
\begin{eqnarray}\small
D^E_{i,k}(v_0(t^m_{0,k}))=\begin{cases} \label{delta}
 \gamma_{i,k} \Big[\frac{1}{2} \Bar{u} \max(v^d_k-v_0(t^m_{0,k}),0)\Big]  \ \ \ \ \ \  i > k\\
\\
0\ \ \ \ \ \  \textnormal{otherwise},
\end{cases}
\end{eqnarray}\\
where $\gamma_{i,k} \in (0,1)$ is the aforementioned discount factor. We model this as a function of the distance $z_{i,k}(t^m_{0,k})$ between vehicles $i$ and $k$ in the merging sequence $k \in \mathcal{K}$. Here, we set $\gamma_{i,k}= \exp(-\beta z_{i,k})$ where $\beta \in (0,1)$ is a parameter.

\textbf{AV Objective Function ($J_{A_k}$) Evaluation:} To evaluate $J_{A_k}$ in (\ref{objective function}) we need to obtain an expression for the AV merging time $t^m_{0,k}$ and energy consumption $E_{0,k}$ in the merging sequence $k$. Since $J_{A_k}$ pertains to the AV objective function when it selects the sequence $k \in \mathcal{K}$, the associated optimal control problem, denoted by $P_k$, is as follows:

\textbf{Problem $P_k$}:
\begin{equation}\label{eqn:tot_car_obj_}
 \begin{aligned}\min_{u_{0}(t),t^m_{0,k}} J_{C_k}(u_0(t),t_{0,k}^m)= \alpha (t^m_{0,k}-t^a_0) + (1-\alpha) \int_{t_0^a}^{t^m_{0,k}} \frac{1}{2}u_0^2(t)dt \end{aligned},
\end{equation}
subject to \eqref{VehicleDynamics}, $x_0(t^m_{0,k})=L_{CZ}$, \eqref{SafeMerging}, \eqref{VehicleConstraints1} and \eqref{VehicleConstraints2}. Note that this formulation does not include any rear-end safety constraint since there is no preceding vehicle for the AV prior to merging.

 The solution of $P_k$ is premised on HDV $k$ allowing the AV to merge ahead of it. We assume that this is the case as long as the AV selects a safe merging time $t^m_{0,k}$ and merging velocity, $v_0(t^m_{0,k})$ to ensure that it positions itself safely between HDVs $k-1$ and $k$ (if they exist). Therefore, we will modify next the safe merging constraint \eqref{SafeMerging} in $P_k$, since \eqref{SafeMerging} only considers HDV $k-1$. 

\textbf{Safe merging time and merging velocity sets:} When both HDVs $k$ and $k-1$ exist, HDV $k$ has to provide space for the AV while the AV also has to satisfy the safe merging constraint \eqref{SafeMerging} with HDV $k-1$. As a result, the AV must not violate either the safety constraint of HDV $k$ or the safe merging constraint \eqref{SafeMerging} of the AV at the same time. By applying \eqref{Safety} and \eqref{SafeMerging}, this implies:
\begin{gather}
z_{0,k-1}(t^m_{0,k}) \geq \varphi_c v_0(t^m_{0,k})+\delta\\
z_{{k},0}(t^m_{0,k}) \geq \varphi_h v_{k}(t^m_{0,k})+\delta . 
\end{gather}
By adding these two constraints, we obtain the following:
\begin{equation} \label{safe_gap}
    z_{k,k-1}(t^m_{0,k}) \geq \varphi_c v_0(t^m_{0,k})+\varphi_h v_{k}(t^m_{0,k})+2\delta,
\end{equation}
where $z_{k,k-1}(t^m_{0,k})$ is the distance between HDVs $k-1$ and $k$ at the merging time, $v_{k}(t^m_{0,k})$ is the velocity of HDV $k$ at the merging time, and $\varphi_h$ is the reaction time for the HDVs (which can be estimated).

Note that the newly derived constraint \eqref{safe_gap} depends on the velocity of HDV $k$ at the merging time, $v_{k}(t^m_{0,k})$ which the AV needs to estimate. Since we assume that the HDVs maintain a constant speed prior to the MP, we set $v_{k}(t^m_{0,k})=v_k^d $. Therefore, the constraint \eqref{safe_gap} can be rewritten as follows:
\begin{equation} \label{safe_gap_conservative}
    z_{k,k-1}(t^m_{0,k}) \geq \varphi_c v_0(t^m_{0,k})+\varphi_h v_k^d+2\delta.
\end{equation}
Since $v_{k}(t^m_{0,k})=v_k^d$, it is easy to derive the positions of HDVs $k$ and $k-1$ at the merging time $t^m_{0,k}$ as follows:
\begin{align} \label{kineticeq}
    x_k(t^m_{0,k}) = x_k(t^a_0) + (t^m_{0,k}-t^a_0)v_k^d, \nonumber \\
    x_{k-1}(t^m_{0,k}) = x_{k-1}(t^a_0) + (t^m_{0,k}-t^a_0)v_{k-1}^d.
\end{align}
By combining \eqref{kineticeq} and \eqref{safe_gap_conservative}, we can now define the following safe sets for the AV merging velocity $v_0(t^m_{0,k})$ and merging time $t^m_{0,k}$:
\begin{equation}\label{mergingvel_set}
    S^v_{0,k} = \Bigl\{v_0(t^m_{0,k}) \in \mathcal{R}^+: v_0(t^m_{0,k}) \leq v^u_{0,k}\Bigl\}
\end{equation}
\begin{equation}\label{mergingtime_set}
    S^t_{0,k} (v_0(t^m_{0,k}))= \Bigl\{t^m_{0,k} \in \mathcal{R}^+: t^l_{0,k}(v_0(t^m_{0,k})) \leq   t^m_{0,k} \leq t^u_{0,k}\Bigl\},
\end{equation}
where $v^u_{0,k} = \frac{1}{\varphi_c} \left(z_{k,k-1}(t^m_{0,k}) -\varphi_h v_k^d -2 \delta \right), \  t^l_{0,k}(v_0(t^m_{0,k})) = t^a_0 + \frac{1}{v_{k-1}^d} \left( x_0(t^m_{0,k})+\varphi_c v_0(t^m_{0,k}) +\delta - x_{k-1}(t^a_0)\right)$, and $t^u_{0,k} =t^a_0 + \frac{1}{v_k^d} \left(x_0(t^m_{0,k}) - \varphi_h v_k^d- \delta - x_{k}(t^a_0)) \right)$.   
Clearly, any $v_0(t^m_{0,k}) \in S^v_{0,k} $ and $t^m_{0,k} \in S^t_{0,k} (v_0(t^m_{0,k}))$ also satisfy \eqref{safe_gap} since they are derived from the more conservative \eqref{safe_gap_conservative}. The satisfaction of \eqref{safe_gap_conservative} implies that there exists a safe gap between HDVs $k-1$ and $k$ and there will be a safe merging time and merging velocity for the AV in merging sequence $k$.

\begin{remark} The constraint \eqref{safe_gap} can be conservative in the sense that it is assumed the HDVs start to decelerate as soon as they observe the AV (see Fig .\ref{fig:merging}). However, the HDVs may in fact start to decelerate earlier when they realize that the AV is about to merge onto their road. It is also possible to create a simple ``signaling scheme'' between an AV an a HDV if we assume the existence of limited communication capabilities between vehicles.
\end{remark}

\textbf{Solution of Problem $P_k$:} We now return to problem $P_k$ with \eqref{SafeMerging} replaced by \eqref{mergingtime_set}, \eqref{mergingvel_set}.
Note that the final states in this problem are constrained by $x_0(t^m_{0,k})=L_{CZ}$ and the merging velocity is chosen from the set $S^v_{0,k}$. Ignoring the vehicle limitations in \eqref{VehicleConstraints1} and \eqref{VehicleConstraints2} (we address this issue in Remark 2), this problem can be readily solved as a fixed final state optimal control problem \cite{bryson2018applied} to give:
\vspace{-2mm}
\begin{align}
    u_0^*(t)=a_0t+b_0 \label{optimal_u}\\
    v_0^*(t)=\frac{1}{2}a_0t^2+b_0t+c_0\label{optimal_v}\\
    x_0^*(t)=\frac{1}{6}a_0t^3+\frac{1}{2}b_0t^2+c_0t+d_0\label{optimal_x},
\end{align}
The four coefficients above  can be computed by using the initial and final conditions of $P_k$, i.e., $x_0(t^a_0)$, $v_0(t^a_0)$, $L_{CZ}$, and $v_0(t^m_{0,k})$. 
The first three are given, but $v_0(t^m_{0,k})$ and $t^m_{0,k}$ are still undetermined and will be derived in the next subsection. Nonetheless, we can solve for these coefficients as functions of $v_0(t^m_{0,k})$ and $t^m_{0,k}$ through the following system of four linear equations of the form $\mathbf{T}_0\mathbf{b}_0=\mathbf{q}_0$ (where, for simplicity, we drop the time arguments and write $x_0(t^a_0)= x_0$, $v_0(t^a_0)=v_0$, $v_0(t^m_{0,k})=v^m_{0,k}$ and, $x_0(t^m_{0,k})=x^m_{0,k}$):
\begin{align}
\begin{bmatrix}
\frac{1}{6}(t^a_0)^3 & \frac{1}{2}(t^a_0)^2 & t^a_0 & 1\\
\frac{1}{2}(t^a_0)^2 & t^a_0 & 1 & 0\\
\frac{1}{6}(t^m_{0,k})^3 & \frac{1}{2}(t^m_{0,k})^2 & t^m_{0,k} & 1\\
\frac{1}{2}(t^m_{0,k})^2 & t^m_{0,k} & 1 & 0
\end{bmatrix}\begin{bmatrix}
a_0\\
b_0\\
c_0\\
d_0\\
\end{bmatrix}=\begin{bmatrix}
x_0\\
v_0\\
L_{CZ}\\
v^m_{0,k}\\
\end{bmatrix}
\end{align}
whose solution is of the form 
\begin{equation} \label{b_0}
    \textbf{b}_0(v_0(t^m_{0,k})),t^m_{0,k})=\mathbf{T_0}^{-1}\mathbf{q_0}.
\end{equation}
emphasizing the fact that the solution still depends on $v_0(t^m_{0,k})$ and $t^m_{0,k}$. 
We can now use $u_0^*(t)$ in (\ref{optimal_u}) with $a_0$, $b_0$ from above to derive the expression for $E_{0,k}$ in (\ref{objective function}) that we have been seeking, as a function of $v_0(t^m_{0,k})$ and $t^m_{0,k}$:
\begin{align} \label{CAV_Energy_AZ}
        E_{0,k}&(v_0(t^m_{0,k})),t^m_{0,k}) =\int_{t_0^a}^{t^m_{0,k}} \frac{1}{2}u_0^2(t)dt =  \int_{t_0^a}^{t^m_{0,k}} \frac{1}{2}(a_0t+b_0)^2dt\\\nonumber
        &=\frac{2(t^m_{0,k}-t_0^a)^2(v_0^2+v_0v^m_{0,k}+v^{m^2}_{0,k})}{(t^m_{0,k}-t_0^a)^3}\\\nonumber
        &+\frac{-6(t^m_{0,k}-t_0^a)(L_{CZ}-x_0)(v_0+v^m_{0,k})}{(t^m_{0,k}-t_0^a)^3} + \frac{6(L_{CZ}-x_0)^2}{(t^m_{0,k}-t_0^a)^3},
\end{align}


\textbf{Solution of problem \eqref{objective function} and optimal index determination:} Now that all terms in \eqref{objective function} have been expressed as functions of $v_0(t^m_{0,k})$ and $t^m_{0,k}$, we proceed to solve it, given $\gamma$ and $\Bar{u}$, as well as $v^d$ and $x_i(t_0^a)$ which are observed/obtained by the AV. We can then obtain a complete solution to the optimal merging problem (\ref{eqn:tot_car_obj}).
This is accomplished in three steps as follows.

\textbf{Step 1:}
Determine the optimal values of $v_0(t^m_{0,k})$ and $t^m_{0,k}$ which minimize the cost $J_k(t^m_{0,k},v^m_{0,k})$ for each $k \in \mathcal{K}$:
\begin{equation} \label{minproblemfixedk}
    \min_{t^m_{0,k},v^m_{0,k}} J_k(t^m_{0,k},v^m_{0,k}), ~~~k \in \mathcal{K}
\end{equation}
subject to \eqref{mergingvel_set}, \eqref{mergingtime_set}, and \eqref{VehicleConstraints1}. The solution provides the optimal cost $J^*_k$ for each $k \in \mathcal{K}$.

\textbf{Step 2:}
Determine the optimal index $k^*$ (equivalently the optimal merging sequence).
This is a simple comparison problem:
\begin{align} \label{argmin problem}
    k^*=\textnormal{argmin}_{k \in \mathcal{K}} J^*_k
\end{align}

\textbf{Step 3:}
Given $k^*$ and $t^{m^*}_{0,k^*}$, $v^{m^*}_{0,k^*}$, the values of $a_0,b_0,c_0,d_0$ are determined through (\ref{b_0}) and provide the complete AV optimal trajectory through 
\eqref{optimal_u},\eqref{optimal_v}, and \eqref{optimal_x}.

\begin{remark}
As already mentioned, in obtaining the solution \eqref{optimal_u},\eqref{optimal_v}, and \eqref{optimal_x} we ignored the vehicle limitations in \eqref{VehicleConstraints1}, \eqref{VehicleConstraints2}. Therefore, it is possible that the AV trajectory determined in Step 3 above is infeasible. There are two ways to address this issue: $(i)$ Simply ignore the index $k^*$ corresponding to this solution and seek the next smallest cost $J^*_k$ with $k \ne k^*$ until a feasible option is found, if any. If none is found, the AV can always wait for all HDVs to get ahead of it, which is always feasible. $(ii)$ Adopt the OCBF method in \cite{XIAO2021} whereby we optimally track the infeasible trajectory subject to constraints employing Control Barrier Functions (CBFs) which guarantee that all constraints (\ref{Safety}), (\ref{SafeMerging}), \eqref{VehicleConstraints1}, \eqref{VehicleConstraints2} are satisfied in such trajectories.
\end{remark}

An analytical solution to problem \eqref{minproblemfixedk} for each $k \in \mathcal{K}$ is difficult to obtain in general due to the complex form of $J_k(t^m_{0,k},v^m_{0,k})$. However, solutions are easily obtained numerically, as long as the number $N$ of HDVs in the CZ is relatively small (in fact, it is always bounded by the length $L_{CZ}$ of the CZ). This limits the number of comparisons required in (\ref{argmin problem}) to the cardinality of the set $\mathcal{K}$.

In what follows, we consider cases in which the optimal index is easily determined by bypassing Step 1 above and show that either $k^*=1$ or $k^*=N+1$; when this holds, the AV can handle the whole group of HDVs as a single ``vehicle'' without the need for full knowledge or any assumptions of human driver behavior. To proceed with this analysis, we require the following additional condition.

\textbf{Assumption 2}: The HDVs observed by the AV at $t_0^a$ move with the same speed $v^d_i(t)=v^d$ and maintain an equal distance from each other $z_{i,i-1}(t)=z$, $ \  \forall t \in [t_0^a,t_0^f],   \ i \in \mathcal{N}-\{0\}$. 

Under Assumption 2, $S^v_{0,k}= S^v_0$ since $v^u_{0,k}=v^u_0$ therefore $v^m_{0,k}=v_m$. Moreover, since the distance between the HDVs and the merging velocity are the same among all $k \in \mathcal{K}$, we have $t^m_{0,k+1}-t^m_{0,k}=\frac{z}{v^d}$ for any two consecutive merging options. Further, Assumption 2 implies that $\gamma_{i,k}=\gamma^{i-k}$ where $\gamma \in (0,1)$. Likewise, all HDVs experience the same disruptions $D^t_{i,k}=\gamma^{i-k}D^t$ and $D^E_{i,k}=\gamma^{i-k}D^E$. 

\subsection{Time-optimal merging sequence}
In this subsection, we consider the case where $\alpha = 1$ in 
\eqref{objective function}, corresponding to the time-optimal merging sequence. However, we show that this sequence remains optimal for values of $\alpha$ that satisfy $\alpha_l \leq \alpha \leq 1$ where
\begin{equation} \label{alpha_l}
    \alpha_l = \frac{E^{max}}{E^{max}+\frac{z}{v^d}+\gamma^{N-1}D^t}, 
\end{equation}
with $E^{max}=\frac{1}{2}((v_{max}-v_0)u_{max}+N(v^d-v_{min})\Bar{u})$. In this case, we show next that the optimal index $k^* \in \mathcal{K}$ becomes relatively simple to determine and is often $k^*=1$.

\begin{theorem}Under Assumptions 1-2 and $\alpha_l < \alpha \leq 1$,

$(i)$: If $v^*_m\geq v^d$, then $k^*=1$.

$(ii)$: If $v^*_m < v^d$, then $k^* = q$, where $q$ is determined from
\begin{equation} \label{timeopt_keyineq}
       \frac{1-\gamma^{N-q+1}}{(1-\gamma)(N-q+1)}v^d D^t   \leq z \leq \frac{\gamma^{N-q+1}(1-\gamma^{q-1})}{(1-\gamma)(q-1)}v^d D^t 
\end{equation}
where $z$ is the distance between HDVs.

\begin{proof} 
We limit the proof to the case $\alpha = 1$. The extension to $\alpha_l \leq \alpha < 1$ can be found in [see Appendix. B].

Case $(i)$: Omitting the energy terms, the optimal value of $J_k$ in \eqref{objective function} becomes:
\begin{equation} \label{alpha_1_obj}
    J^*_k=t^{m^*}_{0,k}-t_0^a+\sum_{i=1}^{N}t^m_{i,N+1}+\sum_{i=k}^{N}\gamma^{i-k} D^t ,
\end{equation}
Since $v^*_m \geq v^d$, based on (\ref{epsilon}), $J_1^*=t^{m^*}_{0,1}-t_0^a+\sum_{i=1}^{N}t^m_{i,N+1}$ 

By its definition, the AV merging time $t^{m^*}_{0,k}$ is monotonically increasing in $k$, therefore, $\min_{k \in \mathcal{K}} t^{m^*}_{0,k}=t^{m^*}_{0,1}$ and it follows that argmin$_{k \in \mathcal{K}} J^*_k=1$.

Case $(ii)$: 
First, consider $k<q$ in (\ref{timeopt_keyineq}). Since (\ref{alpha_1_obj}) holds, by subtracting $J^*_k$ from $J^*_q$ we have:
\begin{align}
    J^*_q-J^*_k &= t^{m^*}_{0,q} - t^{m^*}_{0,k} -\bigl(\gamma^{N-q+1}+...+\gamma^{N-k} \bigl) D^t 
\end{align}
By Assumption 2, we can write $t^{m^*}_{0,q}-t^{m^*}_{0,k}=\frac{(q-k)z}{v^d}$. Since $J^*_q \leq J^*_k$ for all $k \in \{1,...,q-1\}$,
it follows that
\begin{align} \label{z_lowerboundk}
    z \leq \frac{\gamma^{N-q+1}+...+\gamma^{N-k}}{q-k}  v^dD^t = G_1(k),  \ k \in \{1,...,q-1\}
\end{align}
It is easy to show that $G_1(k)$ is monotonically increasing in $k$ for $\gamma < 1$, therefore 
$\min_kG_1(k) = G_1(1)$ for all $k \in \{1,...,q-1\} $. 
As a result, $z \leq G_1(1)$ in (\ref{z_lowerboundk}).

Next, consider $k > q$. 
By subtracting $J^*_q$ from $J^*_k$ we have:
\begin{align}
    J^*_k-J^*_q = t^{m^*}_{0,k} - t^{m^*}_{0,q} -\bigl(\gamma^{N-k+1}+...+\gamma^{N-q} \bigl) D^t 
\end{align}
Using $t^{m^*}_{0,k}-t^{m^*}_{0,q} = \frac{(k-q)z}{v^d}$ and since $J^*_q \leq J^*_k$ for all $k \in \{q+1,...,N+1\}$,
we get:
\begin{align} \label{z_upperboundk}
    z \geq  \frac{\gamma^{N-k+1}+...+\gamma^{N-q}}{k-q} v^d D^t = G_2(k) 
\end{align}
where $G_2(k)$ is monotonically increasing in $k$ for $\gamma < 1$, therefore $\max_kG_2(k) = G_1(N+1)$ for all $k \in \{q+1,...,N+1\}$. Therefore, $z \geq G_2(N+1)$. 
By combining the two cases above, \eqref{timeopt_keyineq} is obtained.
\end{proof}
\end{theorem}
\begin{remark}
Theorem 1 requires knowledge of $v_m^*$ which is obtained from solving problem \eqref{minproblemfixedk}. This can be avoided by selecting instead the highest merging velocity from the set $S^v_0$, since only time optimality is of interest and we also know that the speed will be the same for all possible merging sequences under Assumption 2.
\end{remark}

\subsection{Energy-optimal merging sequence}
\begin{figure}
\vspace{-5mm}
\centering
\includegraphics[scale=0.4]{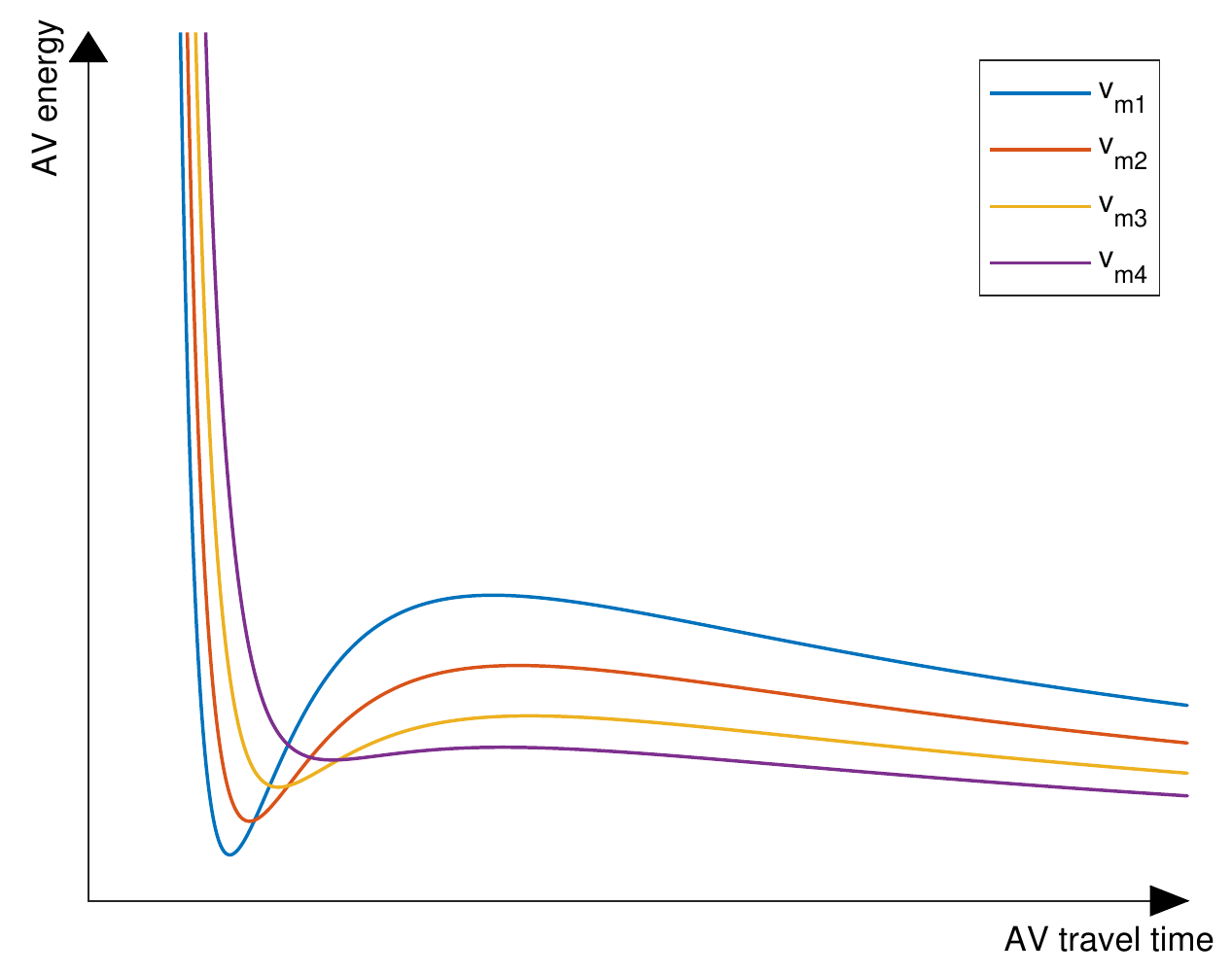} 
\caption{AV energy function with different merging velocities}
\label{fig:CAV_E_diff_vm}
\end{figure}

We now consider the case where $\alpha = 0$ in \eqref{objective function}. 
Similar to the time-optimal case of the previous section, it can be shown that the energy-optimal sequence is unchanged for all $\alpha$ such that
$0 \leq \alpha \leq \alpha_u$ where, 
\begin{equation} \label{alpha_u}
    \alpha_u = \frac{E^{min}}{E^{min}+\frac{Nz}{v^d}+\frac{1-\gamma^N}{1-\gamma}D^t},
\end{equation}
 and $E^{min} = \gamma ^{N-1} D^E$.
 We show next that under easy-to-check conditions the optimal index is either $k^*=1$ or $k^*=N+1$, i.e., the AV either merges ahead of or behind the entire HDV group. Before proceeding, let us take a closer look at the AV energy function derived in \eqref{CAV_Energy_AZ}, which depends on the AV's travel time and distance and its initial and merging velocities. Under Assumption 2, however, the expression in \eqref{CAV_Energy_AZ} depends, for all $k \in \mathcal{K}$, only on the AV travel time. 
Thus, the energy expression in \eqref{CAV_Energy_AZ} is a rational function of the AV's travel time, $t_k = t^m_{0,k}-t_0^a $ which takes the general form: 
\begin{equation} \label{CAV_Energy_function_form}
    E_0(t_k)=\frac{A_1t_k^2+A_2t_k+A_3}{t_k^3}, \ \ \forall k \in \mathcal{K},
\end{equation} 
where $A_1=2(v_0^2+v_0v_m+v^2_m)>0$, $A_2=-6(L_{CZ}-x_0)(v_0+v_m)<0$ and $A_3=6(L_{CZ}-x_0)>0$ are all constants. 
It can be shown that this function has both vertical and horizontal asymptotes at $ E_{0}(t_k)=0$ and $ E_{0}(t_k)=\infty$ (see Fig. \ref{fig:CAV_E_diff_vm}). The following lemma establishes a property of this function to be used in Theorem 2 (the proof can be found in [see Appendix. C]).

\textbf{Lemma 1}: The function $E_{0}(t_k)$ in \eqref{CAV_Energy_function_form} is monotonically decreasing in $k\in \mathcal{K}$ only if  $ t_{{N+1}} \leq  \tau_1$,
where
\begin{equation} \label{tau_1}
\tau_1=\frac{-A_2-\sqrt{A_2^2-3A_1A_3}}{A_1}.
\end{equation}

\begin{theorem} Under Assumptions 1-2 and $0 \leq \alpha \leq \alpha_u$, if
\begin{equation} \label{timethreshold}
    t^{m^*}_{0,N+1} - t_0^a  \leq \frac{3(L_{CZ}-x_0)}{v_0+v^*_m+\sqrt{v_0v^*_m}}
\end{equation}
\begin{equation}
       \text{then,} ~~~~ \min J^*_k = \min (J^*_1,J^*_{N+1})
\end{equation}


\vspace{+3mm}
\begin{proof}
We limit the proof to the case $\alpha = 0$. The extension to $0 < \alpha \leq \alpha_u$ can be found in [see Appendix.D].
We will consider two cases:

$(i)$  k = 1:
By setting $\alpha = 0$ and $k = 1$ in \eqref{objective function} and \eqref{delta}, the objective function can be written as: $J^*_1 = E_{0,1}$

$(ii)$ $ 2 \leq k \leq N+1 $:
Under Assumption 2, by setting $\alpha = 0$ in the objective function in \eqref{objective function} we can rewrite the minimization problem as follows:
\begin{equation} \label{obj_alpah_0}
    \min_k J^*_k = \min_k\left( E_{0,k}+ \left(\sum_{i=k}^{N}\gamma^{i-k} D^E \right)  \right).
\end{equation}
If the minimum of the first term and second term occur at the same $k$, the minimization of the summation becomes the summation of the minimization.
The second term is monotonically decreasing in $k$ since, under Assumption 2 for $2 \leq k \leq N+1$, $v_m = v^m_{0,k}$ in \eqref{delta} therefore $\textnormal{argmin}_k \left(\sum_{i=k}^{N}\gamma^{i-k} D^E \right) = N+1$.
As for the first term,
according to Lemma 1, for $E_{0}(t_k)$ in \eqref{CAV_Energy_function_form} to be monotonically decreasing in $k$, $t_{N+1}\leq \tau_1 $ needs to be satisfied, where 
$\tau_1=\frac{3(L_{CZ}-x_0)}{v_0+v^*_m+\sqrt{v_0v^*_m}}$
using the values of $A_1, A_2, A_3$ with $v_m = v_m^*$.
Since $t_k \leq t_{N+1}$ for all $k \in \mathcal{K}-\{1\}$, 
we get $E_{0}(t_k) > E_{0}(t_{k+1})$ for all $k \in \mathcal{K}-\{1\}$,
which leads to $\textnormal{argmin}_k E_{0,k} = N+1$.

Since $k = N+1$ minimizes both terms in \eqref{obj_alpah_0}, we can write $k^* = N+1$. Finally, 
the minimum of $J_k^*$ for every $k \in \{1,...,N+1\}$ is given by
$\min J^*_k = \min (J^*_1,J^*_{N+1})$.
\end{proof}
\end{theorem}

Note that in Theorem 1, we managed to obtain a condition on $z$ for the optimality over all merging sequences, whereas in Theorem 2 we are only able to derive a sufficient condition dependent on the optimal AV travel time when $k=N+1$ not exceeding a fixed threshold $\tau_1$ given by (\ref{timethreshold}). In this case, 
the AV can ignore all but the first and last merging sequence, since either of these two is optimal. 
This is due to the non-convexity of the energy as a function of travel time which makes it difficult to derive optimality conditions for all merging sequences, as seen in \ref{fig:CAV_E_diff_vm} showing the AV energy with respect to travel time under different merging velocities, $v_{m_1} > v_{m_2} >v_{m_3} > v_{m_4} > 0$.

\section{SIMULATION RESULTS} \label{sec:simulation}
All simulations of the merging problem shown in Fig. \ref{fig:merging} 
are based on \textsc{PTV Vissim} and all algorithms are
implemented using \textsc{MATLAB} and \textsc{ode45} to integrate the AV dynamics.
The parameters used are: $L_{CZ} = 400\textnormal{m}, \varphi_c = 1.8\textnormal{s},\varphi_h = 1.5\textnormal{s}, \delta = 3.78 \textnormal{m}, \Bar{u}=2, \gamma = 0.4, \beta= 0.1, u_{max} = 4.905 \textnormal{m/s}^2, u_{min} = -5.886\textnormal{m/s}^2, v_{max} = 30\textnormal{m/s},t_0^a = 0, v_{min} = 0\textnormal{m/s}, v^d=16.66,v_0=16.66$. The simulations have been performed for a variety of initial positions of the AV and HDVs to be able to thoroughly validate the optimality results we have derived. 

\begin{figure}
\centering
\includegraphics[scale=0.47]{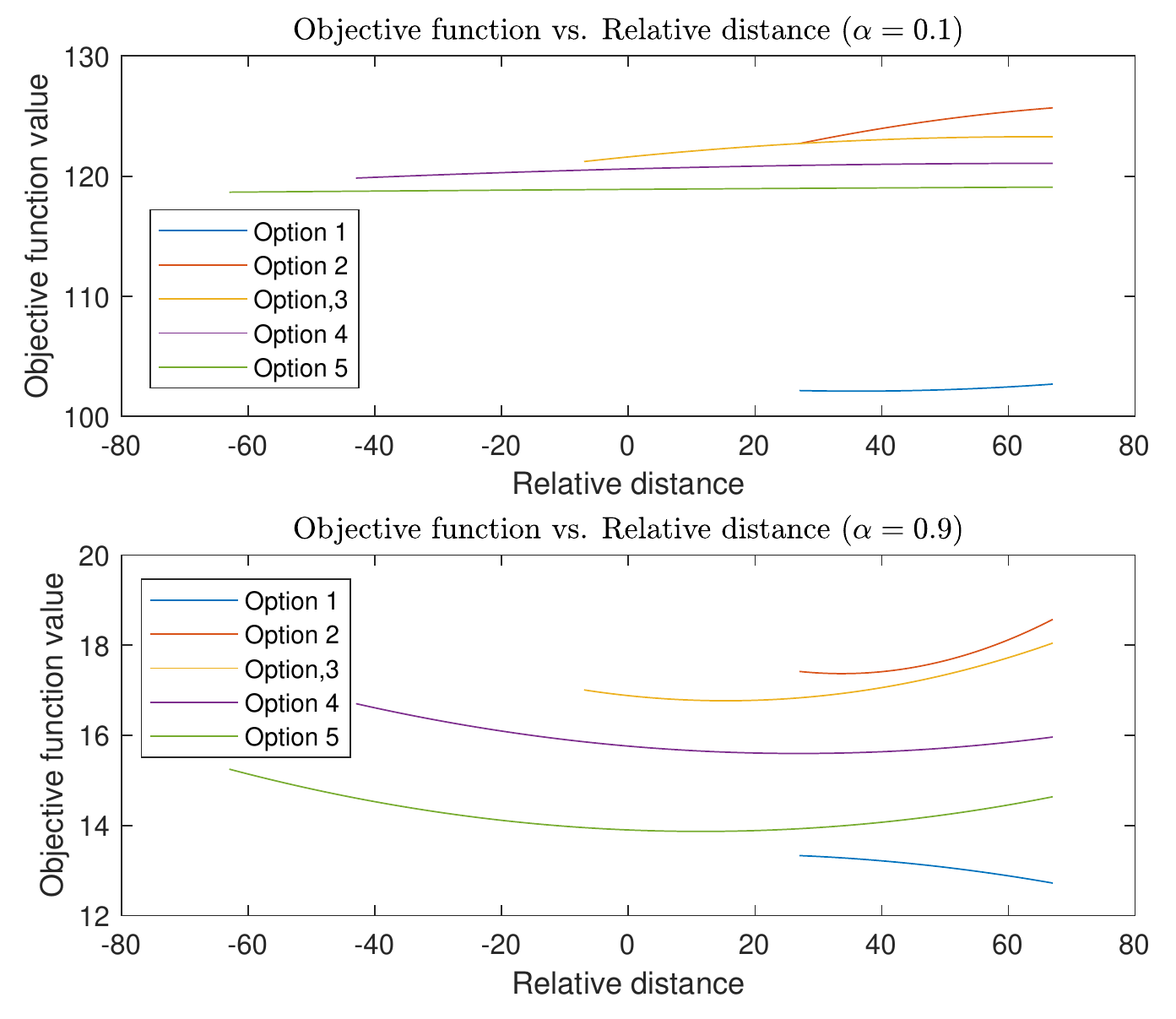} 
\caption{Objective function value with respect to relative distance ($z = 22$)}
\label{fig:dx_22}
\end{figure}

\begin{figure}
\centering
\includegraphics[scale=0.4]{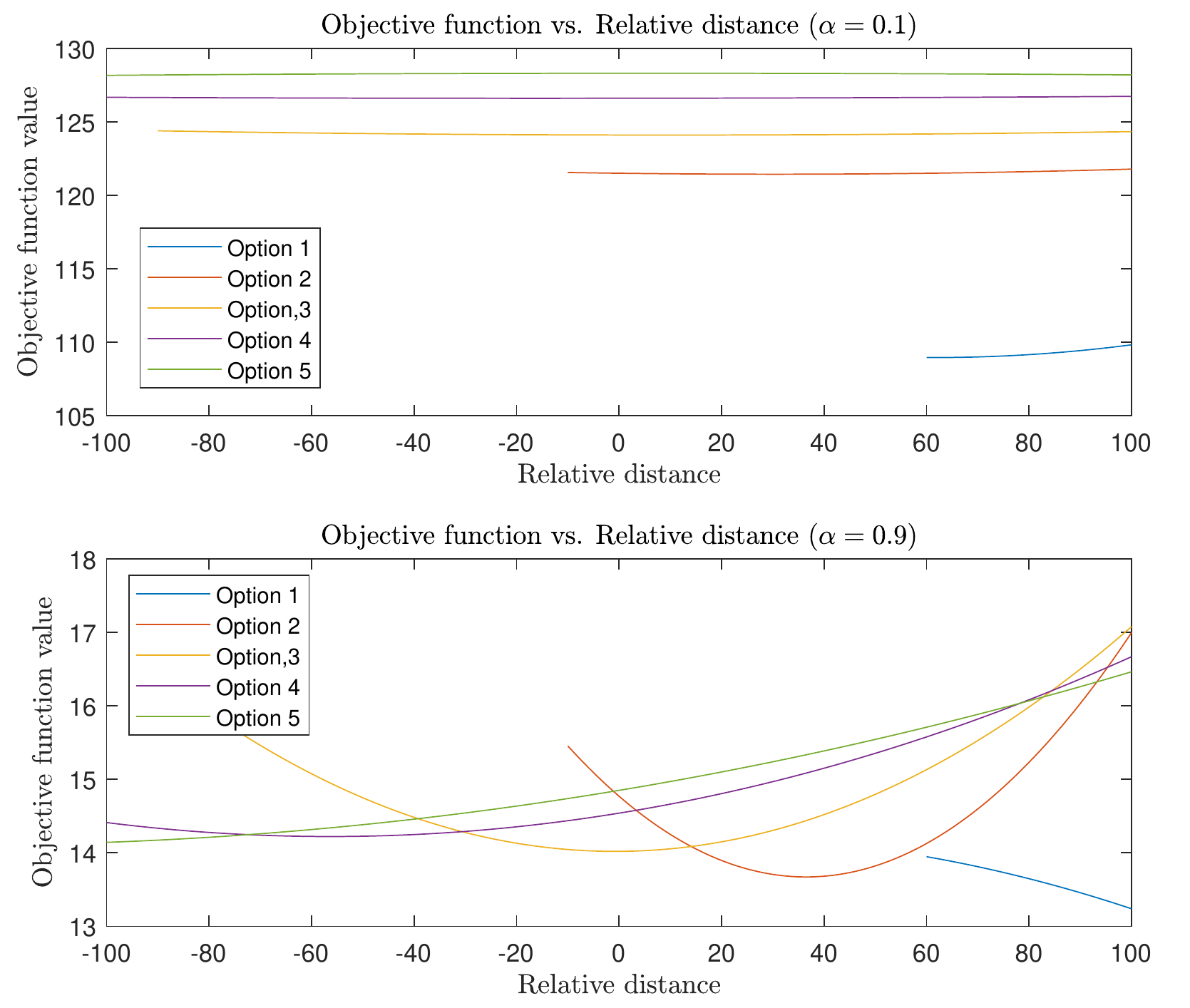} 
\caption{Objective function value with respect to relative distance ($z = 40$)}
\label{fig:dx_40}
\end{figure}

\begin{table*}
        \centering
        \begin{tabular}{|c|c|c|c|c|c|c|c|c|c|c|c|}
            \hline
            HDV initial position & $x_0(t_0^a)$ & \multicolumn{2}{c|}{$J_1$} & \multicolumn{2}{c|}{$J_2$} & \multicolumn{2}{c|}{$J_3$}& \multicolumn{2}{c|}{$J_4$}& \multicolumn{2}{c|}{$J_5$} \\
        \cline{3-12}  
        & & $\alpha = 0.1$&$\alpha = 0.9$ &$\alpha = 0.1$&$\alpha = 0.9$ &$\alpha = 0.1$&$\alpha = 0.9$& $\alpha = 0.1$&$\alpha = 0.9$ &$\alpha = 0.1$&$\alpha = 0.9$\\
        \hline
        \multirow{4}{*}{\makecell{\\$x_1(t_0^a)=330$ \\ $x_2(t_0^a)=308$ \\ $x_3(t_0^a)=286$ \\ $x_4(t_0^a)=264$} } & 234 & inf & inf & inf & inf & inf & inf & inf & inf & \textcolor{red}{118.71} & \textcolor{blue}{15.25}\\
        \cline{2-12}
        & 254 & inf & inf & inf & inf  & inf & inf & 119.49 & 16.76 & \textcolor{red}{118.72} & \textcolor{blue}{14.60} \\
        \cline{2-12}
        & 274 & inf & inf & inf & inf  & inf &  inf & 120.18 & 16.08 & \textcolor{red}{118.81} & \textcolor{blue}{14.16} \\
        \cline{2-12}
        & 294 & inf & inf & inf & inf  & 121.35 & 17.0 & 120.51 & 15.80 & \textcolor{red}{118.88} & \textcolor{blue}{13.92} \\
        \cline{2-12}
        & 314 & inf & inf & inf & inf & 122.28 & 15.56  & 120.74 & 15.53 & \textcolor{red}{119.0} & \textcolor{blue}{ 13.88} \\
        \cline{2-12}
        & 334 & \textcolor{red}{102.09} & \textcolor{blue}{13.26} & 123.47 & 17.25 & 123.0 & 17.04 & 121.02 & 15.66 & 119.02 & 14.02 \\
        \cline{2-12}
        & 364 & \textcolor{red}{102.41} & \textcolor{blue}{12.92} & 125.53 & 18.21 & 123.16 & 17.69 & 121.11 & 16.07 &119.06 & 14.38   \\
        \hline
        \multirow{4}{*}{\makecell{  \\ $ $ \\$x_1(t_0^a)=330$ \\ $x_2(t_0^a)=290$ \\ $x_3(t_0^a)=250$ \\ $x_4(t_0^a)=210$ } }  & 200 & inf & inf & inf  & inf &\textcolor{red}{ 124.30} & 15.52 & 126.65 & \textcolor{blue}{14.25} & 128.25 & 14.26\\
        \cline{2-12}
        & 220 & inf & inf & inf  & inf  & \textcolor{red}{124.21} & 14.72 & 126.64 & \textcolor{blue}{14.16} & 128.25 &14.33\\
        \cline{2-12}
        & 240 & inf & inf & \textcolor{red}{121.62}  & 17.45 & 124.15 & \textcolor{blue}{14.15} & 126.55 & 14.16 & 128.27 & 14.47\\
        \cline{2-12}
        & 260 & inf & inf  & \textcolor{red}{121.58}  & 15.43 & 124.12 & \textcolor{blue}{13.90} & 126.57 & 14.39 & 128.29 & 14.67\\
        \cline{2-12}
        & 280 & inf & inf & \textcolor{red}{121.53}  & 14.22 & 124.09 & \textcolor{blue}{14.01} & 126.59 & 14.59 & 128.32 & 14.94 \\
        \cline{2-12}
        & 300 & inf & inf & \textcolor{red}{121.46}  & \textcolor{blue}{13.58} & 124.17 & 14.41 & 126.68 & 14.97 & 128.36 & 15.28 \\
        \cline{2-12}
        & 320  & inf & inf & \textcolor{red}{121.48}  & \textcolor{blue}{13.80} & 124.18 & 14.87 & 126.69 & 15.49 & 128.40 & 15.68 \\
        \cline{2-12}
        & 340 & \textcolor{red}{109.0} & \textcolor{blue}{13.82} & 121.59 & 14.74 & 124.28 & 15.74 & 126.77 & 16.13 & 128.26 & 15.98 \\
        \cline{2-12}
        & 360 & \textcolor{red}{109.44} & \textcolor{blue}{13.44} & 121.67 & 16.25 & 124.15 & 16.55 & 126.51 & 16.26 & 128.17 & 16.01 \\
        \hline
        
        \end{tabular}
        \caption{Objective function values of merging sequences corresponding to sufficiently large and sufficiently small $\alpha$. }
        \label{Table I}
\end{table*} 
\textbf{Time-optimal and energy-optimal cases}: Table \ref{Table I} summarizes the results of 16 simulations for 4 HDVs and 1 AV performing the merging process. To validate Theorems 1 and 2, the objective value corresponding to each option has been provided in the table for sufficiently small $\alpha = 0.1$ and large $\alpha = 0.9$ (costs corresponding to the optimal options are shown in red and blue, respectively). In the table, "inf" indicates the infeasibility of a particular merging sequence due to the AV's limitations in terms of maximum/minimum acceleration and speed. The table data are visually depicted in Figs. \ref{fig:dx_22} and \ref{fig:dx_40} (for inter-HDV distances $z=22$ and $z=40$ respectively) showing the objective value as a function of the relative distance between the HDV group and the AV for all sequences. The relative distance is measured from the projection of the AV onto the other road to the center of the whole HDV group: $x_0^\prime(t_0^a) = x_0(t_0^a)-\frac{x_1(t_0^a)+x_r(t_0^a)}{2}$. 

Theorem 1 implies that the time-optimal merging sequence may be independent of the AV's initial position relative to the HDVs, leading to $k^*=1$. This can be seen in the figures, as long as this sequence is feasible. When this is not feasible, the AV selects the optimal sequence depending on the relative distance. Theorem 2 is also illustrated for energy-optimal sequences when $z$ is sufficiently small so that a disruption would be too costly, therefore the energy-optimal index is either $k=1$ or $k=5$.

\begin{figure}
\centering
\includegraphics[scale=0.4]{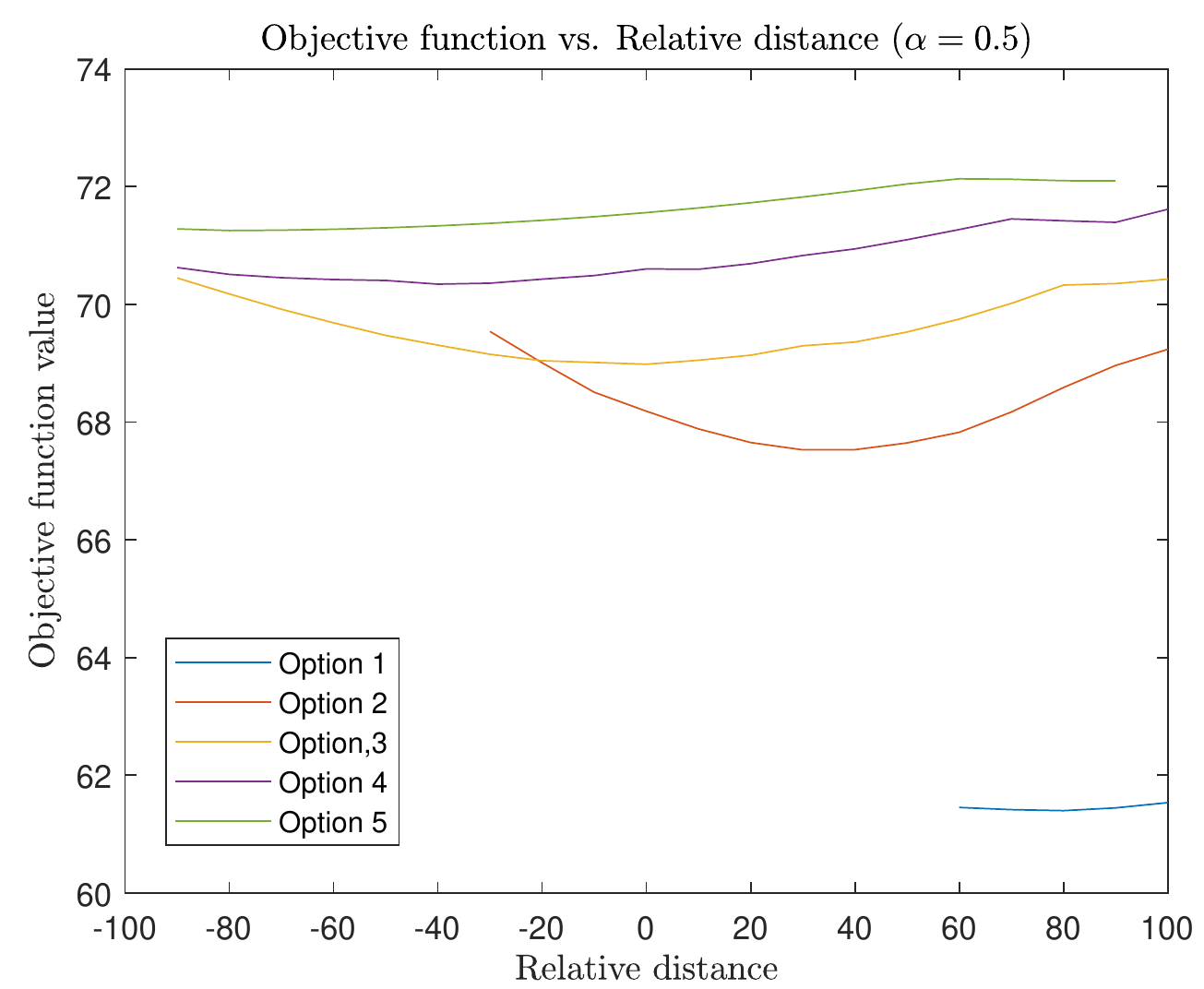} 
\caption{Objective function value with respect to relative distance ($z = 40$)}
\label{fig:dx_40_alpha_0.5}
\end{figure}


\textbf{General case}: When neither Theorem 1 or 2 applies, we need to use a numerical method to calculate the optimal merging sequence through (\ref{minproblemfixedk}) and (\ref{argmin problem}). Thus we choose $\alpha = 0.5$ so that neither time or energy optimality dominates. 
As seen in Fig. \eqref{fig:dx_40_alpha_0.5} for $ z = 40$, different values of the index $k$ are optimal depending on the relative distance shown.

Finally, in terms of computational complexity, the 3-steps solution approach in \eqref{minproblemfixedk} and \eqref{argmin problem} takes approximately $50$ msec to evaluate each option, therefore a total of approximately $50 \times N$ msec is needed for all possible merging sequences.

\section{CONCLUSIONS} \label{sec:conclude}

We have studied the merging problem that the integration of a single AV with a group of HDVs as a potential ``building block'' for more complex mixed traffic situations involving a relatively small number of CAVs (since AVs may now communicate among them) interacting with HDVs. An index optimal policy is derived to determine a joint time and energy-optimal  merging sequence for the AV so as to benefit the whole system.


\section{Appendix}
A.\begin{figure}
\centering
\includegraphics[scale=0.4]{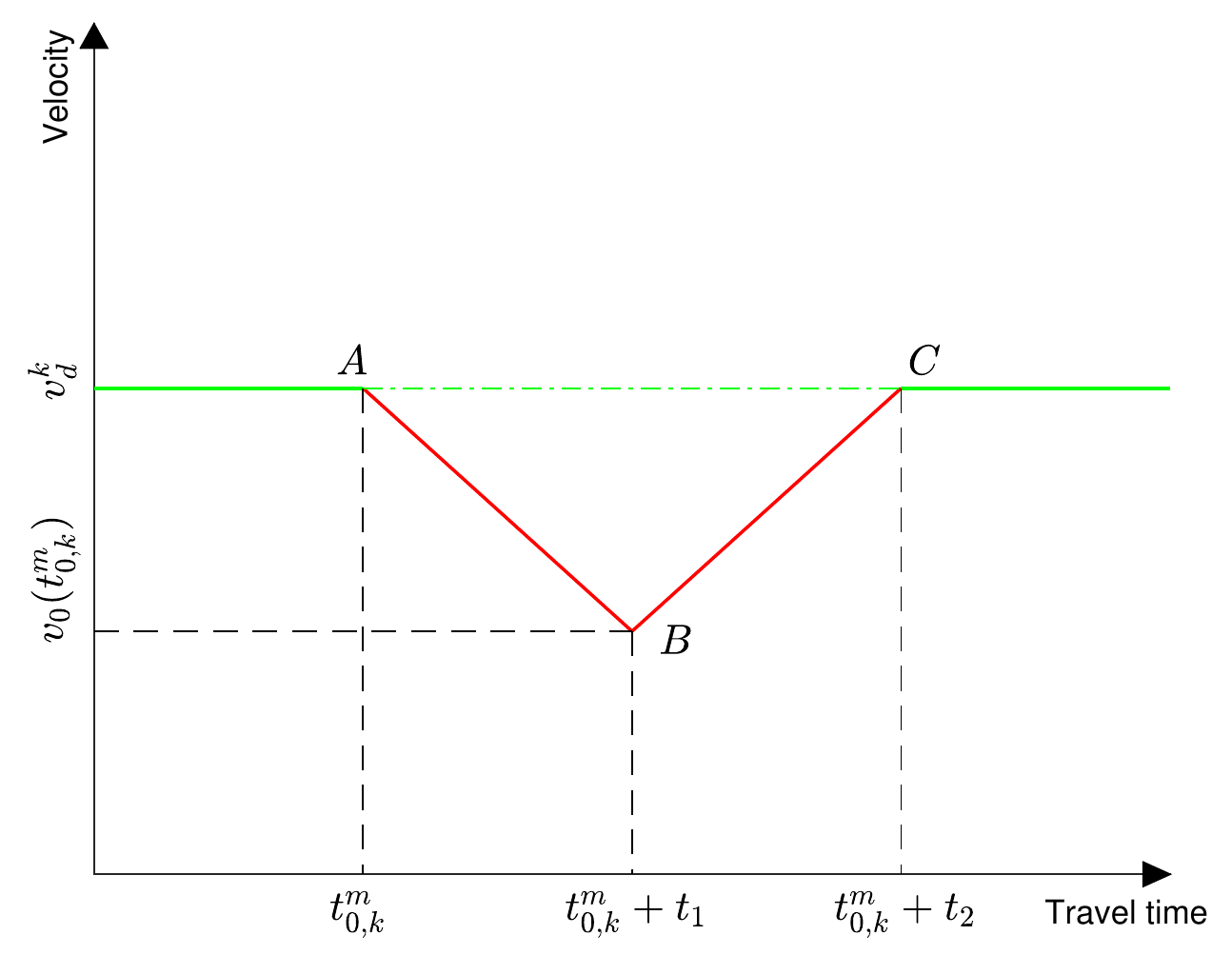} 
\caption{The estimated velocity of HDV $k$ in merging sequence $k$ (case ($i$))}
\label{fig:HDV i velocity}
\end{figure}
To obtain the expression for the $D^t_{k,k}$ and $D^E_{k,k}$ we consider two cases: (i) When the merging velocity of the AV is less than the steady state velocity of the HDV $k$, $v_0(t^m_{0,k}) < v_k^d $. In this case, the HDV has to decelerate which is assumed to be at a constant deceleration rate $\Bar{u}$. The amount of deceleration will be proportional to the difference between these two speeds hence it is safe to assume that the HDV has to decelerate until at least it reaches the velocity of the AV at the merging. Once it reaches that velocity it will start to accelerate again since both HDV and AV are in the AL. Fig. \ref{fig:HDV i velocity} can illustrate the velocity trajectory of the HDV $k$ which can provide an approximate measure of the energy consumption and travel time delay of the HDV $k$. As we know the area under the velocity-time graph gives the total displacement of the object. As a result, the area of triangle $\triangle{ABC}$ in the Fig. \ref{fig:HDV i velocity} illustrates the additional distance HDV $k$ would have traveled before time $t^m_{0,k}+t_2$ if the AV had not merged before it (i.e. due to disruption and slowing down the total traveled distance by HDV $k$ will be less than when it maintains its steady speed). For the travel time delay, we need to calculate the triangle area denoted by $ S_{ABC}$, which is equal to the distance that HDV is lagging behind the no-disruption case.
\begin{equation}
    S_{ABC} = \frac{t_2 (v_k^d-v_0(t^m_{0,k}))}{2}.
\end{equation}
Then, the delay in HDV travel time will be the triangle area divided by the steady-state velocity as follows:
\begin{equation}
     D^t_{k,k}=\frac{ S_{ABC}}{v^d_k} = \frac{t_2 (v_k^d-v_0(t^m_{0,k}))}{2v_k^d},
\end{equation}
where due to symmetry, $t_2=2t_1$ and $t_1=\frac{v_k^d-v_0(t^m_{0,k})}{\Bar{u}}$. As a result, $D^t_{k,k}$ can be rewritten as:
\begin{equation}
    D^t_{k,k}=\frac{(v_k^d-v_0(t^m_{0,k}))^2}{v_k^d\Bar{u}}.
\end{equation}
By using the energy metric $\frac{1}{2}u_k^2$ for HDV $k$ we can approximate the energy that HDV $k$ spend as follows:
\begin{equation}
    D^E_{k,k}=\frac{1}{2}\Bar{u}^2 t_1+\frac{1}{2}\Bar{u}^2 (t_2-t_1).
\end{equation}
After replacing $t_1$ and $t_2$, the final expression for the energy loss of the HDV $k$ can be rewritten as:
\begin{equation}
    D^E_{k,k}=\Bar{u}(v_k^d-v_0(t^m_{0,k})).
\end{equation}
(ii). When the merging velocity of the AV is greater than the steady state velocity of the HDV $k$, $v_0(t^m_{0,k}) > v_k^d $. In this case, there will be no impact on the HDV as the AV can safely merge with higher velocity than HDV $k$. Therefore
\begin{align}
    D^t_{k,k}= 0,\\
    D^E_{k,k}= 0.
\end{align}
In order to represent these two cases in a compact form we will use the $\max$ function as follows:
\begin{gather}
    D^t_{k,k}=\frac{\max(v_k^d-v_0(t^m_{0,k}),0)^2}{\Bar{u}v_k^d},\nonumber\\
    D^E_{k,k} = \Bar{u} \max(v_k^d-v_0(t^m_{0,k}),0).
\end{gather}
Since we are solving a minimization problem up until the merging point, due to symmetry the travel time delay and energy loss up until the merging point can be written as follows:
\begin{gather}
    D^t_{k,k}=\frac{\max(v_k^d-v_0(t^m_{0,k}),0)^2}{2\Bar{u}v_k^d}\nonumber\\
    D^E_{k,k} = \frac{1}{2}\Bar{u} \max(v_k^d-v_0(t^m_{0,k}),0)
\end{gather}

\begin{figure}
\vspace{-5mm}
\centering
\includegraphics[scale=0.43]{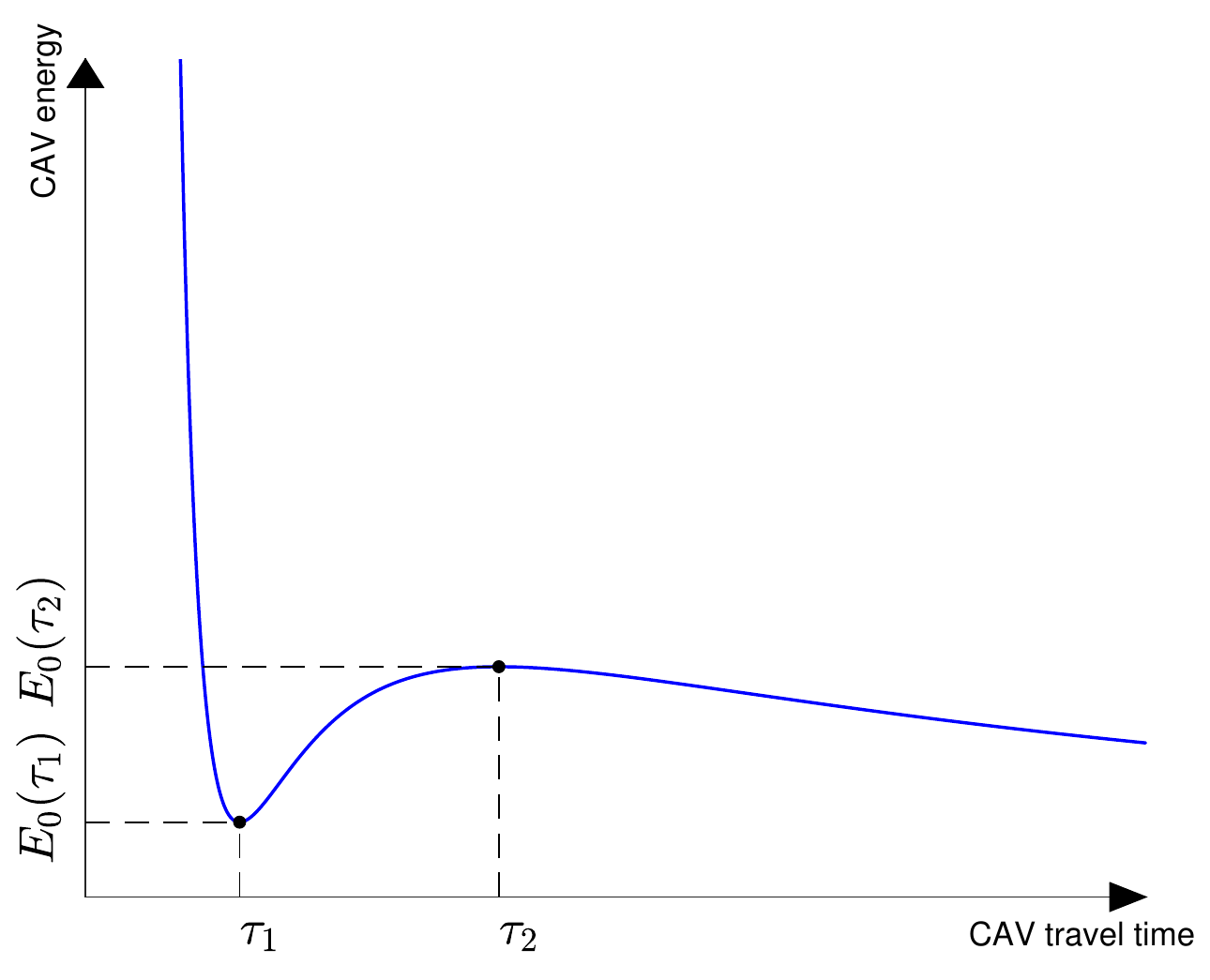} 
\caption{AV energy function}
\label{fig:CAV_Energy_function}
\end{figure}

B. In this section, we intend to prove that the theorem 1 results hold for a range of $\alpha$. Instead of proving each case, we consider a general case as follows. If we consider $J_k$ in \eqref{objective function} as a function of $\alpha$, $J_k(\alpha)$, let $k^*$ be the minimizer of $J^*_k(1)$ in \eqref{argmin problem} (i.e. in case $(i)$ is 1 and in case $(ii)$ is $q$). In the following we will propose a range of $\alpha$ within which $k^*$ remains optimal, 
\begin{equation} \label{eq.B.1}
    k^* = \textnormal{argmin}J^*_k(\alpha), \ \forall \alpha \in (\alpha_l,1].
\end{equation}
Where $\alpha_l$ defined in \eqref{alpha_l}.
For \eqref{eq.B.1} to hold we need to have: $J^*_{k^*} \leq J^*_k, \ \forall k \in \mathcal{K}$. Based on \eqref{objective function} we can write:
\begin{align}
    \alpha \left( t^{m^*}_{0,k^*}-t_0^a+\sum_{i=1}^{N}t^{m^*}_{i,k^*} \right)   +(1-\alpha) \left(E_{0,k^*}+ \sum_{i=1}^{N}E_{i,k^*}  \right) \leq \nonumber \\ 
    \alpha \left( t^{m^*}_{0,k}-t_0^a+\sum_{i=1}^{N}t^{m^*}_{i,k} \right)   +(1-\alpha) \left(E_{0,k}+ \sum_{i=1}^{N}E_{i,k}  \right).
\end{align}
By simplifying and solving it for $\alpha$ it results:
\begin{equation}\label{eq.B.3}
    \alpha \geq \frac{\sum_{i=0}^{N}(E_{i,k^*}-E_{i,k})}{\sum_{i=0}^{N}(t^{m^*}_{i,k} -t^{m^*}_{i,k^*})+\sum_{i=0}^{N}(E_{i,k^*}-E_{i,k})} := F_1(k,k^*).
\end{equation}
To ease up the notation we simply write $E_{k^*-k}=\sum_{i=0}^{N}(E_{i,k^*}-E_{i,k})$ and $t_{k-k^*}=\sum_{i=0}^{N}(t^{m^*}_{i,k} -t^{m^*}_{i,k^*})$. In order for \eqref{eq.B.3} to hold $ \forall k,k^* \in \mathcal{K}$ we can write:
\begin{equation}\label{eq.B.4}
    \alpha \geq \max_{k,k^*} F_1(k,k^*).
\end{equation}
 Obtaining the maximum value of $F_1(k,k^*)$ can be challenging therefore we decide to derive an upper bound, $\alpha_l$ instead. To do so, instead of treating $F_1$ as a function of $k,k^*$, we consider the whole terms $E_{k^*-k}$ and $t_{k-k^*}$ as arguments such that $F_1(E_{k^*-k},t_{k-k^*})$. By taking derivative of $F_1$ with respect to $E_{k^*-k}$ and $ t_{k-k^*}$, it can be shown that $F_1$ is increasing in $E_{k^*-k}$ and decreasing in $t_{k-k^*}$ therefore $\alpha_l$ can be proposed as the upper bound for $F_1(k,k^*)$ as follows:
\begin{align} \label{app_alpha_l}
  \max_{k,k^*} F_1(k,k^*) \leq \frac{\max_{k,k^*}E_{k^*-k}}{\min_{k,k^*}t_{k-k^*}+\max_{k,k^*}E_{k^*-k}}:=\alpha_l. 
\end{align}
By defining $\alpha_l$ as in \eqref{app_alpha_l}, for $\alpha \geq \alpha_l$, ultimately \eqref{eq.B.1} is satisfied. 
\begin{equation}
    \alpha \geq \alpha_l \geq \max_{k,k^*} F_1(k,k^*).
\end{equation}
An expression for $\alpha_l$ yet remains to be determined.
To evaluate $\alpha_l$ we need to calculate $\max_{k,k^*}E_{k^*-k}$ and $\min_{k,k^*}t_{k-k^*}$ which roughly speaking will translate into, the maximum difference in the energy consumption of any $k,k^*  \in \mathcal{K}$, and the minimum difference in the travel time of any $k,k^*  \in \mathcal{K} $, respectively.\\
Let us start with $\min_{k,k^*}t_{k-k^*}$. Based on the definition of $\min_{k,k^*}t_{k-k^*}$ and under assumptions 1-2 we can write:
\begin{align} \label{eq.B.6}
   \min_{k,k^*}t_{k-k^*} &=  \min_{k,k^*} t^{m^*}_{0,k}-t^{m^*}_{0,k^*}+\sum_{i=1}^{N}t^{m^*}_{i,k} -t^{m^*}_{i,k^*}\\ \nonumber
   & = \min_{k,k^*} \frac{(k-k^*)z}{v^d}+\bigl(\gamma^{N-k+1}+...+\gamma^{N-k^*} \bigl) D^t.
\end{align}
It can be seen that the minimzer of the $\min_{k,k^*}t_{k-k^*}$ happens at $k^* = 1$ and $k = 2$. Since $k$ and $k^*$ are integers therefore $\min_{k,k^*}k-k^*=1$. Note that since $k^*$ is time optimal merging sequence we know that $k > k^*$.To calculate the minimizer of the second term in \eqref{eq.B.6} since $\gamma <1$ therefore $k^* = 1$ will minimize the second term and as a result $k=k^*+1=2$. Consequently, the following expression can be written for $\min_{k,k^*}t_{k-k^*}$.
\begin{equation}\label{tkk*}
    \min_{k,k^*}t_{k-k^*} = \frac{z}{v^d}+\gamma^{N-1}D^t.
\end{equation}
Moving on to the next term in $\alpha_l$ defined in \eqref{app_alpha_l}, $\max_{k,k^*}E_{k^*-k}$. To obtain an expression we use the fact that the maximum of summation is always less than or equal to the summation of the maximum. Then by separating AV and HDVs energy consumption based on \eqref{objective function} we have:
\begin{align}\label{eq.B.8}
   \max_{k,k^*}  E_{k^*-k} \leq \max_{k,k^*} E_{0,k^*}-E_{0,k} + \max_{k,k^*} \sum_{i=1}^{N}D^E_{i,k^*}-D^E_{i,k}.
\end{align} 
Since $k^*,k \in \mathcal{K}$ can be basically any two unidentical merging sequences, to maximize each term in \eqref{eq.B.8}, we can consider $k^*$,  a merging sequence with the maximum energy consumption for all cars and $k$ to be a merging sequence with total energy consumption of $0$. Namely, in merging sequence $k^*$, AV accelerates with $u_{max}$ in \eqref{VehicleConstraints2}, until it reaches $v_{max}$ in \eqref{VehicleConstraints1}, and whole group of HDVs will decelerate with $\Bar{u}$ until they reach $v^*_m = v_{min}$ in \eqref{VehicleConstraints1} which lead to maximum disruption, $D^E_{i,k}$ defined in \eqref{delta}.
Now based on \eqref{energy} and \eqref{delta} we can write the following expression for $\max_{k,k^*}E_{k^*-k}$ as follows:
\begin{align} \label{Ek*k}
   \max_{k,k^*} E_{k^*-k} & \leq \max_{k,k^*} E_{0,k^*}-E_{0,k} + \max_{k,k^*} \sum_{i=1}^{N}D^E_{i,k^*}-D^E_{i,k} \\ \nonumber
   &=\frac{1}{2}(((v_{max}-v_0)u_{max}-0)+(N(v^d-v_{min})\Bar{u}-0)).
\end{align}
By replacing \eqref{tkk*} and \eqref{Ek*k} into \eqref{app_alpha_l} it results:
\begin{equation}
    \alpha_l= \frac{\frac{1}{2}((v_{max}-v_0)u_{max}+N(v^d-v_{min})\Bar{u})}{\frac{1}{2}((v_{max}-v_0)u_{max}+N(v^d-v_{min})\Bar{u})+\frac{z}{v^d}+\gamma^{N-1}D^t}
\end{equation}
C. The energy function defined in \eqref{CAV_Energy_function_form}, is a rational function with constant coefficients $\forall k \in \mathcal{K}$.
\begin{equation}
    E_0(t_k)=\frac{A_1t_k^2+A_2t_k+A_3}{t_k^3}.
\end{equation}
To analyze the behavior of this function, we consider its derivative. The equation $E^{\prime}_0(t_k) =0$, where $E^{\prime}_0(t_k)$ is the derivative of the function with respect to $t_k$, has two zeros. As a result, this function has two extremums: one minimum, $E_0(\tau_1)$, and one maximum $E_0(\tau_2)$, see Fig \ref{fig:CAV_Energy_function}.  
\begin{align}
    E^{\prime}_0(t_k) &= \frac{A^{\prime}_1t_k^2+A^{\prime}_2t_k+A^{\prime}_3}{t_k^4}\\ \nonumber
    \tau_1 &= \frac{-A_2-\sqrt{A_2^2-3A_1A_3}}{A_1} , \tau_2 = \frac{-A_2+\sqrt{A_2^2-3A_1A_3}}{A_1}
\end{align}
where $A^{\prime}_1 = -A_1,A^{\prime}_2=-2A_2,A^{\prime}_3 = -3A_3 $. Since $A^{\prime}_1 <0$, $A^{\prime}_2 > 0$ and $A^{\prime}_3 < 0$ are all constants, we can easily determine the sign of the $E^{\prime}_0(t_k)$ as follows: $E^{\prime}_0(t_k) < 0,  \ \forall t_k \in (0,\tau_1)$, $E^{\prime}_0(t_k) > 0,  \ \forall t_k \in (\tau_1,\tau_2)$ and finally $E^{\prime}_0(t_k) < 0 \ \forall t_k \in (\tau_2,\infty)$. Therefore the function $ E_0(t_k)$ is monotonically decreasing over the interval $(0,\tau_1]$.
Now if we  define the energy function of the AV in the time interval of $[t_{1},t_{N+1}]$, if $t_{N+1} < \tau_1$ the energy function is monotonically decreasing in $t_k$ or equivalently in $k \in \mathcal{K}$.\\

D. Likewise part B, if $k^*$ is the minimizer of $J^*_k(0)$, in some cases, we are able to propose a range of $\alpha$ within which $k^*$ remains optimal.
\begin{equation} \label{eq.C.1}
    k^* = \textnormal{argmin}J^*_k(\alpha), \ \forall \alpha \in [0,\alpha_u),
\end{equation}
where $\alpha_u$ defined in \eqref{alpha_u}.
For \eqref{eq.C.1} to hold we need to have: $J^*_{k^*} \leq J^*_k, \ \forall k,k^* \in \mathcal{K}$. Based on \eqref{objective function} we can write:
\begin{align}
    \alpha \left( t^{m^*}_{0,k^*}-t_0^a+\sum_{i=1}^{N}t^{m^*}_{i,k^*} \right)   +(1-\alpha) \left(E_{0,k^*}+ \sum_{i=1}^{N}E_{i,k^*}  \right) \leq \nonumber \\ 
    \alpha \left( t^{m^*}_{0,k}-t_0^a+\sum_{i=1}^{N}t^{m^*}_{i,k} \right)   +(1-\alpha) \left(E_{0,k}+ \sum_{i=1}^{N}E_{i,k}  \right).
\end{align}
By simplifying and solving it for $\alpha$ it results:
\begin{equation}\label{eq.C.3}
    \alpha \leq \frac{\sum_{i=0}^{N}(E_{i,k}-E_{i,k^*})}{\sum_{i=0}^{N}(t^{m^*}_{i,k^*} -t^{m^*}_{i,k})+\sum_{i=0}^{N}(E_{i,k}-E_{i,k^*})} := F_2(k,k^*).
\end{equation}
To ease up the notation we simply write $E_{k-k^*}=\sum_{i=0}^{N}(E_{i,k}-E_{i,k^*})$ and $t_{k^*-k}=\sum_{i=0}^{N}(t^{m^*}_{i,k^*} -t^{m^*}_{i,k})$. In order for \eqref{eq.C.3} to hold $ \forall k,k^* \in \mathcal{K},$ we can write:
\begin{equation}\label{eq.C.4}
    \alpha \leq \min_{k,k^*} F_2(k,k^*).
\end{equation}
 Obtaining the minimum value of $F_2(k,k^*)$ can be challenging therefore we decide to derive a lower bound, $\alpha_u$ instead. Similar to the analysis of part B, $\alpha_u$ can be proposed as the lower bound for $F_2(k,k^*)$ as follows:
\begin{align} \label{app_alpha_u}
  \min_{k,k^*} F_2(k,k^*) \geq \frac{\min_{k,k^*}E_{k-k^*}}{\max_{k,k^*}t_{k^*-k}+\min_{k,k^*}E_{k-k^*}}:=\alpha_u.
\end{align}
By defining $\alpha_u$ as in \eqref{app_alpha_u}, for $\alpha \leq \alpha_u$, ultimately \eqref{eq.C.1} is satisfied. 
\begin{equation} \label{eq.C.7}
    \alpha \leq \alpha_u \leq \min_{k,k^*} F_2(k,k^*),
\end{equation}
the expression for $\alpha_u$ yet remains to be determined.To evaluate $\alpha_u$, we can follow the same procedure as the previous section by calculating $\max_{k,k^*}t_{k^*-k}$ and $\min_{k,k^*}E_{k-k^*}$.\\
Let us start with $\max_{k,k^*}t_{k^*-k}$. Based on the definition of $\max_kt_{k^*-k}$ and under Assumptions 1-2 we can write:
\begin{align} \label{tkk*}
   \max_{k,k^*}t_{k^*-k} &=  \max_{k,k^*} t^{m^*}_{0,k^*}-t^{m^*}_{0,k}+\sum_{i=1}^{N}t^{m^*}_{i,k^*} -t^{m^*}_{i,k}\\ \nonumber
   & = \max_{k,k^*} \frac{(k^*-k)z}{v^d}+\bigl(\gamma^{N-k^*+1}+...+\gamma^{N-k} \bigl) D^t.
\end{align}
Similar to the former analysis it can be seen that the maximizer of the $\max_{k,k^*}t_{k^*-k}$ happens at $k^* = N+1$ and $k = 1$. 
\begin{equation}
    \max_{k,k^*}t_{k^*-k} = \frac{Nz}{v^d}+\frac{1-\gamma^N}{1-\gamma}D^t.
\end{equation}
Moving on to the next term in $\alpha_u$ defined in \eqref{app_alpha_u}, $\min_{k,k^*}E_{k-k^*}$. Following the same procedure by separating AV and HDVs energy consumption, based on \eqref{objective function} we can write:
\begin{align}\label{eq.C.8}
   \min_{k,k^*}  E_{k-k^*} \leq \min_{k,k^*} E_{0,k}-E_{0,k^*} + \min_{k,k^*} \sum_{i=1}^{N}D^E_{i,k}-D^E_{i,k^*}.
\end{align} 
To solve for \eqref{eq.C.8}, in contrast to what we did in part B, we need to consider two cases:
\begin{enumerate}
    \item $v_m^* \geq v^d$ (without disruption case):
    Since $k, k^* \in \mathcal{K}$ can be any merging sequence with arbitrary conditions, in this case, one can provide two merging sequences that are not identical yet optimal with the same energy consumption but different total travel times, therefore, in this case, $\alpha$ can only be zero. Due to the fact that the AV energy function is not a one-to-one function with respect to travel time, it is quite possible that two different merging sequences which obviously differ in total travel time will consume the same energy. (see. Fig \ref{fig:CAV_Energy_function} and Fig \ref{fig:CAV_E_diff_vm}). As a result in this case there is not such a range for $\alpha$. Namely,
    \begin{align}\label{Ekk*_case1}
   \min_{k,k^*} E_{k-k^*} & \leq \min_{k,k^*} E_{0,k}-E_{0,k^*} + \min_{k,k^*} \sum_{i=1}^{N}D^E_{i,k}-D^E_{i,k^*} =0.
\end{align}
    As a result, From \eqref{tkk*} and \eqref{Ekk*_case1}, \eqref{eq.C.7} results:
    \begin{equation}
        \alpha \leq \alpha_u \leq 0.
    \end{equation}
    \item $v_m^* < v^d$ (with disruption): In this case, although it is still possible for AV to consume the same energy in two different merging sequences the disruption in HDVs will make a difference in terms of merging sequence energy consumption.
\begin{align}\label{Ekk*_case2}
   \min_{k,k^*} E_{k-k^*} & \leq \min_{k,k^*} E_{0,k}-E_{0,k^*} + \min_{k,k^*} \sum_{i=1}^{N}D^E_{i,k}-D^E_{i,k^*} \\ \nonumber
   &= \min_{k,k^*}\bigl(\gamma^{N-k^*+1}+...+\gamma^{N-k} \bigl) D^E = \gamma ^{N-1} D^E.
\end{align}
\end{enumerate}
Finally, as a result, From \eqref{tkk*} and \eqref{Ekk*_case2}, \eqref{app_alpha_u} results in :
\begin{equation}
    \alpha_u= \frac{\gamma ^{N-1} D^E}{\gamma ^{N-1} D^E+\frac{Nz}{v^d}+\frac{1-\gamma^N}{1-\gamma}D^t}.
\end{equation}




\end{document}